\documentclass[showpacs, showkeys,preprint, english,11pt]{revtex4}
\usepackage{graphicx}
\usepackage{mathrsfs}
\usepackage[hypertex]{hyperref}
\usepackage{amssymb,amsmath}

\begin{document}

\title{A quintessence scalar field with a constant-equation-of-state potential}
\author{\textsc{Hoavo      Hova}\footnote{hovhoavhill@yahoo.fr}}

\affiliation{D\'{e}partement de physique, Facult\'{e} des Sciences et Techniques,\\ Universit\'{e} de Kara, Togo.}

\begin{abstract}
In this paper we study a quintessence scalar field in a potential exponentially decreasing with e-folding number, and show that the universe is currently undergoing an accelerated expansion and could probably remain accelerating indefinitely.
\end{abstract}


\maketitle

\section{Introduction}
The recent accelerated expansion, observed in the dynamics of the universe \cite{Riess, Perlmutter, Tegmark, Seljak, Adelman, Abazajian, Spergel1, Page, Hinshaw, Jarosik}, is regarded as a result of a currently dominant energy fluid, dubbed ``\emph{dark energy}'',  with an equation of state $\omega<-\frac{1}{3}$. In particle physics dark energy is interpreted  as quantum vacuum energy \cite{Weinberg}, modeled by the cosmological constant $\Lambda$ of general relativity, for which $\omega_{\Lambda} = -1$. The density of this vacuum energy is uniform, and constant throughout the universe, invariable with time. This form introduced by Einstein is compatible with all astronomical data and is the only model not introducing new degrees of freedom. If dark energy does take this form, this means that it is a fundamental property of the universe. Unfortunately, the observed value of the cosmological constant is much less than its value computed from quantum field theory. The cosmological constant thus faces the fine-tuning and coincidence puzzles \cite{Amendola1}. 

Owing to the lack of the knowledge of origin and nature of such  an energy component with negative pressure, one may add some dynamical contribution to the total energy-momentum tensor to have an effective equation of state $\omega_{eff}<-\frac{1}{3}$ for the expansion to be accelerated. Hence, a dynamic, time-evolving and spatially dependent form of energy with negative pressure is sufficient to explain the observed accelerating universe, and also to alleviate both fine-tuning and coincidence puzzles faced by the cosmological constant. Subsequently, many models of dark energy have been proposed (see the review  article \cite{Amendola1} and references therein), among which \textit{quintessence} \cite{Ratra, Ratra1, Zlatev, Caldwell, Wetterich, Wetterich1, Coble, Turner, Boyle, Saridakis} is considered as describing a scalar field $\varphi$ minimally coupled to gravity and with an equation of state that changes with time in the range $-1\le\omega_{\varphi}\le 1$. In contrast to the cosmological constant that is a non-dynamical potential energy in an absolute vacuum, devoid of matter (CDM, baryonic) or radiation, an accelerated expansion in quintessence, however, is mostly controlled by a variable scalar-field potential $V(\varphi)$ \cite{Copeland}. Hence, several potentials, such as exponential potential \cite{Copeland1, Ng}, power-law type potential \cite{Ratra, Caldwell}, etc., have been investigated to lead successfully to an accelerating universe.

In this paper, we are going to work with a time-evolving equation of state of a quintessence scalar field with a potential, depending directly on the e-folding number  and computed by considering a constant equation of state for the quintessence field. We show that such a potential could be responsible for the late time accelerated expansion. Some analytical solutions are straightforwardly found, while those approximated are determined under certain assumptions.

In Section II, we consider a constant equation of state and derive the corresponding scalar field potential. In Section III we investigate the dynamics of the universe with a time-evolving equation of state. In Section IV  will be presented a summary of the results and the conclusions. Throughout this paper an overdot denotes differentiation with respect to the time coordinate  $t$, and we use units in which $c=\hbar=\kappa^{2}=1$. 

\section{A constant equation of state}

 We are to determine the potential of a quintessence scalar field with a constant equation of state (EoS). 
The EoS $\omega_{\psi}$ of a quintessence scalar field $\psi$ with a potential $V(\psi)$ is defined by 
\begin{equation}
\omega_{\psi}\equiv \dfrac{p_{\psi}}{\rho_{\psi}}=\dfrac{\dot{\psi}^{2}-2 V(\psi)}{\dot{\psi}^{2} + 2 V(\psi)}.
\end{equation}
Assume the EoS $\omega_{\psi}$ remains constant during the evolution of the universe and equals $\omega_{c}$ such that $ -1 < \omega_{c} < -1/3$ to favor an accelerated expansion of the universe at the present stage, one will have
\begin{equation}
\label{a0}
\dot{\psi}^{2}=\dfrac{2(1+\omega_{c})}{1-\omega_{c}}~V(\psi), 
\end{equation}
leading thus to
\begin{equation}
\label{a}
\rho_{\psi}=\dfrac{2}{1-\omega_{c}}~V(\psi) \qquad \mbox{and} \qquad p_{\psi}=\dfrac{2\omega_{c}}{1-\omega_{c}}~V(\psi). 
\end{equation}
Therefore, using the e-folding number  $x=\ln a$ (where $a(t)$ is the scalar factor of an expanding universe) the energy conservation equation, $\dot{\rho}_{\psi}+3H(p_{\psi}+\rho_{\psi})=0$, translates into 
\begin{equation}
\label{a.1}
\dfrac{dV(x)}{dx}+3(1+\omega_{c})V(x) = 0,
\end{equation}
which  gives after integration ($V_{0}=V(x=0)$)
\begin{equation}
\label{b}
V(x)=V_{0} \mbox{e}^{-3 \lambda x}
\end{equation}
with
\begin{equation}
\label{c}
\omega_{c}=-1+\lambda
\end{equation}
for a positive $\lambda$. Acceleration occurs for $\lambda<2/3$.

To reconstruct $V(\psi)$ we shall proceed as follows. The Hubble parameter is given by:
\begin{equation}
\label{c1}
H^{2}=H_{0}^{2}\big[\Omega^{0}_{m}\mbox{e}^{-3x}+(1-\Omega^{0}_{m})\mbox{e}^{-3(1+\omega_{c})x}\big],
\end{equation} 
where we used  
\begin{equation}
V_{0}=\dfrac{3}{2}H_{0}^{2}(1-\omega_{c})(1-\Omega^{0}_{m}) \qquad \mbox{with} \qquad \Omega^{0}_{m}=\dfrac{\rho^{0}_{m}}{3H_{0}^{2}}.
\end{equation}
Using (\ref{a0}), (\ref{b}) and (\ref{c1}) the scalar field $\psi(x)$ reads
\begin{equation}
\label{c2}
\psi(x)=\tilde{\psi}+\dfrac{2}{\sqrt{3}}\dfrac{\sqrt{1+\omega_{c}}}{|\omega_{c}|}\mbox{arcsinh}\bigg[\sqrt{\frac{1-\Omega^{0}_{m}}{\Omega^{0}_{m}}}\exp\big(-\frac{3}{2}\omega_{c}x\big)\bigg],
\end{equation}
$\tilde{\psi}$ being a constant. Finally, eliminating $x$ between (\ref{b}) and (\ref{c2}) and making the change $\psi-\tilde{\psi}\to \psi$  the potential $V(\psi)$ is \cite{Varun}:
\begin{equation}
V(\psi)=\dfrac{3}{2}H_{0}^{2}(1-\omega_{c})(1-\Omega^{0}_{m})\bigg[\dfrac{\Omega^{0}_{m}}{1-\Omega^{0}_{m}}\sinh^{2}\bigg(\dfrac{\psi}{\psi_{0}}\bigg)\bigg]^{1+1/\omega_{c}}
\end{equation}
with $\psi_{0}=\dfrac{2}{\sqrt{3}}\dfrac{\sqrt{1+\omega_{c}}}{|\omega_{c}|}$.
~~~~~~~~~~~~~~~~~~~~~~~~~~~~~~~~~~\\

In the following we are going to study the evolution of the universe  for a time-varying EoS with  the potential $V(x)$ determined in (\ref{b}).

\section{A time--evolving equation of state}

Now we are to consider a minimally coupled quintessence scalar field $\varphi$ in a potential $V(\varphi)$, with energy density $\rho=\dfrac{1}{2}\dot{\varphi}^{2}+V(\varphi)$ and pressure $p=\dfrac{1}{2}\dot{\varphi}^{2}-V(\varphi)$. The  scalar-field EoS $\omega\equiv p/\rho$ may be written in the form
\begin{equation}
\omega=-1+\dfrac{\dot{\varphi}^{2}}{\rho}.
\end{equation} 
In simplifying the EoS  we introduce a new scalar field, let us say $\phi$, related to $\varphi$ via their first derivatives with respect to time $t$ as: 
\begin{equation}
\label{ch}
\dot{\phi}\equiv\dfrac{\varepsilon\dot{\varphi}}{\sqrt{\rho}} \qquad \mbox{with} \qquad \varepsilon=\pm 1.
\end{equation}
Contrary to $\varphi$ that is dimensionless in  units $c=\hbar=\kappa^{2}=1$, the new field $\phi$ has dimension of the inverse of the Hubble parameter $H$,   $\left[\phi \right]=\left[H\right]^{-1} $. Without loss of understanding the new field $\phi$ will be referred to as ``\emph{n-field}'' throughout the rest of the work. 
From (\ref{ch}), knowing the explicit expression of $\rho(\varphi)$  we can express the \emph{n-field} $\phi$ as a function of $\varphi$, given by
\begin{equation}
\label{ch1}
\phi= \tilde{\phi} + \varepsilon \int \dfrac{d\varphi}{\sqrt{\rho(\varphi)}}.
\end{equation}
where $\tilde{\phi}$ is a constant of integration, that will be set to zero from now on.

Using the \emph{n-field} $\phi$ the EoS of the scalar field takes then the tachyonic form \cite{Padmanabhan, Bagla}
\begin{equation}
\omega=-1+\dot{\phi}^{2}.
\end{equation}
Subsequently, the quintessence field behaves as a tachyon field  for $\phi$ varying in the range $0 \le \dot{\phi}^{2} < 1 $. 
On the other hand, expressing the potential in terms of $\dot{\phi}$ leads to
\begin{equation}
V=\dfrac{1}{2}(2-\dot{\phi}^{2})\rho.
\end{equation}
In order to simplify furthermore the calculations we shall now define a new (positive) dimensionless  quantity $\mu$ by
\begin{eqnarray}
\label{ch3}
\mu &\equiv&\dfrac{V}{\rho}\label{1}\\
   &=& 1-\dfrac{1}{2}\dot{\phi}^{2},
\end{eqnarray}
whence we can compute the time-dependence of the new field,
\begin{equation}
\phi(t)=\bar{\phi} + \varepsilon \sqrt{2} \int (1-\mu(t))^{1/2} dt,
\end{equation}
with $\bar{\phi}$ being a constant of integration.
The EoS $\omega$ of the scalar field  is therefore recast in the form
\begin{equation}
\omega= 1 -2\mu.
\end{equation} 
For a tachyon field, $0 \le \dot{\phi}^{2} < 1 \Longrightarrow 1/2 \le \mu < 1$, while a quintessence scalar field leads to $0 \le \dot{\phi}^{2} \le 2 \Longrightarrow 0 \le \mu \le 1$.

The Friedmann equations in an expanding universe dominated by a matter field with a constant EoS $\omega_{m}$  and the quintessence field are 
\begin{eqnarray}
3H^{2}=\rho_{m}^{0}a^{-3(1+\omega_{m})} + \rho \label{ch5}\\
2\dot{H}+ 3H^{2}= -(1-2\mu)\rho -\omega_{m}\rho_{m}^{0}a^{-3(1+\omega_{m})} \label{ch6},
\end{eqnarray}
where $H=\dot{a}/a$ is the Hubble parameter  and $\rho_{m}^{0}$ is the present matter energy density. Eliminating $\rho$ between (\ref{ch5}) and (\ref{ch6}) yields
\begin{equation}
\label{ch.}
\frac{d E^{2}}{dx} + 6(1-\mu)E^{2}=3\Omega^{0}_{m}(1-2\mu -\omega_{m})\mbox{e}^{-3 (1+\omega_{m}) x}
\end{equation}
or
\begin{equation}
\label{ch4.2}
\frac{d E^{2}}{dx} + 3(1+\omega)E^{2}=3\Omega^{0}_{m}(\omega-\omega_{m})\mbox{e}^{-3 (1+\omega_{m}) x}
\end{equation}
where $\Omega^{0}_{m}=\rho_{m}^{0}/3H_{0}^2$ and $E=H/H_{0}$ is the fractional Hubble parameter, with $H_{0} = H(x=0)$. Notice that, when we deal with more than one matter species with constant EoS and no interactions, we have just to sum over the index $m$ in the second hand of Eqs. (\ref{ch.}) or (\ref{ch4.2}).

Furthermore, assuming that the different energy components (such as matter field and scalar field) conserve separately, and using (\ref{ch3}) the conservation equation of the quintessence scalar field, $\dot{\rho}+3H(p+\rho) = 0$, reads
\begin{equation}
\label{ch4}
\dfrac{\dot{V}}{V} - \dfrac{\dot{\mu}}{\mu} + 6 (1-\mu) \dfrac{\dot{a}}{a} = 0 .
\end{equation}
In terms of $x$, (\ref{ch4}) becomes
\begin{equation}
\label{ch4.1}
\dfrac{1}{\mu}\dfrac{d\mu}{dx} - 6 (1-\mu)  = \dfrac{1}{V}\dfrac{d V}{dx} .
\end{equation}
Equation (\ref{ch4.1}) enables one to determine $\mu(x)$ for a given potential $V(x)$ and vice versa. When $\mu = \mu_{0}$ is a constant this equation yields (\ref{a.1}). 

Now we want to consider a potential, $V(x)$, depending upon the e-folding number $x$ and in the following express it in terms of the scalar field $\varphi$ and the \emph{n-field} $\phi$. In view of Eq. (\ref{ch4.1}), the simplest variable potential in integrating easily (\ref{ch4.1}) is an exponential potential in the form $V(x) \propto \exp(\alpha x)$ (with $\alpha$ a constant)  since the corresponding term in (\ref{ch4.1}) will lead to the constant $\alpha$. Without loss of generality we set $\alpha = -3\lambda$ and the potential thus corresponds to the potential (\ref{b}) derived by considering a \emph{constant equation of state} $\omega_{c}$. We write therefore
\begin{equation}
\label{2}
V(x) = V_{0} \mbox{e}^{-3\lambda x},
\end{equation}
where $\lambda$ is a \textit{positive} constant in order to have a decreasing potential during the evolution of the universe up to now. Consider henceforth  a pressureless matter (CDM, baryonic) with $\omega_{m} = 0$, solutions to Eqs. (\ref{ch4.1}) and (\ref{ch4.2}) thus give, respectively  
\begin{equation}
\mu(x)=\dfrac{1-\lambda/2}{1+u\exp(-6(1-\lambda/2)x)}
\end{equation}
and
\begin{equation}
\label{3}
E^{2}(x)=\Omega^{0}_{m}\mbox{e}^{-3 x} + \dfrac{(1-\Omega^{0}_{m})u}{1+u}\mbox{e}^{-6 x} + \dfrac{1-\Omega^{0}_{m}}{1+u}\mbox{e}^{-3\lambda x},
\end{equation}
where we used the initial condition $E(0)=1$, and $u$ is a \emph{positive} constant of integration. $\mu$ being positive, $\lambda$ must then satisfy $\lambda \le 2$ (we will see later that $\lambda \ne 1$ for existence of some solutions). Using (\ref{1}), (\ref{2}) and (\ref{3}) the integration constant $u$ reads
\begin{equation}
\label{4}
u=(1-\lambda/2)(1-\Omega^{0}_{m}) v^{-1} - 1
\end{equation}
with $v=V_{0}/3H^{2}_{0}$. The Hubble parameter therefore reduces to
\begin{equation}
E^{2}(x)=\Omega^{0}_{m}\mbox{e}^{-3x} + (1-\Omega^{0}_{m}-v (1-\lambda/2)^{-1})\mbox{e}^{-6x} + v (1-\lambda/2)^{-1}\mbox{e}^{-3\lambda x}.
\end{equation}
 The EoS of the scalar field, given by
\begin{equation}
\label{5}
\omega(x)=1-\dfrac{2-\lambda}{1+u\exp(-6(1-\lambda/2)x)}
\end{equation}
varies from $\omega_{-\infty}=1$ to $\omega_{+\infty}=-1+\lambda$, showing that the universe could indefinitely stay accelerating in the future if $\omega_{+\infty}<-1/3$, that is, for $\lambda<2/3$. Notice furthermore that $\omega_{+\infty}=-1+\lambda$ corresponds exactly to $\omega_{c}=-1+\lambda$ in  (\ref{c}). Substituting (\ref{4}) into (\ref{5}) the present value $\omega_{0}$ is evaluated as
 \begin{equation}
\label{eq}
\omega_{0}=1-\dfrac{2v}{1-\Omega^{0}_{m}}. 
\end{equation}
 We then see that $\omega_{0}$ is independent of the parameter $\lambda$ and  acceleration occurs  for $v>2(1-\Omega^{0}_{m})/3$. From the Friedmann equation (\ref{ch5}) we have at the present stage
\begin{equation}\label{eq1}
v=1-\Omega^{0}_{m}-\dfrac{\dot{\varphi}_{0}^{2}}{6H_{0}^{2}},
\end{equation}
where $\dot{\varphi}_{0}=\dot{\varphi}(x=0)$ is the present velocity of the scalar field. The parameter $v$ is therefore less than $\Omega^{0}=1-\Omega^{0}_{m}$ for a nonzero $\dot{\varphi}_{0}$, and consequently $\omega_{0}>-1$. Indeed, using (\ref{eq1}) in (\ref{eq}) one obtains
\begin{equation}
\omega_{0}=-1+\dfrac{r}{1-\Omega^{0}_{m}}, 
\end{equation}
where $r=\dot{\varphi}_{0}^{2}/3H_{0}^{2}<1-\Omega^{0}_{m}$.  An accelerating universe thus requires a small scalar-field velocity $r$, then a large value for the present potential $v$: $r<2(1-\Omega^{0}_{m})/3<v$. It is important to mention that, when the scalar-field velocity vanishes at $x=0$ (which means $r=0$),  $v=\Omega^{0}=1-\Omega^{0}_{m}$ and $\omega_{0}=-1$. But this condition will yield a \emph{negative} $u=-\lambda/2$ and $\omega(x)$  will have a singularity at
 $x_{s} =\ln \big(\lambda/2\big)\big/\big[3(2-\lambda)\big] $. So we do consider a nonzero scalar-field velocity $r > 0$ in the present work.

We are going now to compute the $x$-dependence of the scalar field. Using $\rho = \frac{1}{2}\dot{\varphi}^{2} + V$ the scalar field may be determined through the differential equation 
\begin{equation}
\label{6}
\dfrac{d\varphi}{dx}=\varepsilon \bigg[\dfrac{-\sigma'(x)}{\gamma(x) + \sigma(x)}\bigg]^{1/2}, 
\end{equation}
where prime denotes differentiation with respect to $x$, and
\begin{equation}
\label{8}
 \sigma(x) =\dfrac{\rho}{3H_{0}^{2}} = (1-\Omega^{0}_{m}-v(1-\lambda/2)^{-1})\mbox{e}^{-6x} + v (1-\lambda/2)^{-1}\mbox{e}^{-3\lambda x} \quad \mbox{and} \quad \gamma(x)=\dfrac{\rho_{m}}{3H_{0}^{2}}= \Omega^{0}_{m}\mbox{e}^{-3x}.
\end{equation}
We are not able to solve analytically the equation (\ref{6}). However, assume $\lambda$ is of order of unity, we are to consider two cases, on the one hand the epoch dominated by the scalar-field kinetic term only, and  on the other hand the epoch where it is negligible compared to matter and potential terms.

\subsection{Universe dominated by the scalar-field kinetic term}

In early times of the universe ($x\to -\infty$ and $\omega \sim 1$), when the kinetic term of the scalar field (proportional to $\mbox{e}^{-6x}$), dominated over other terms, the scalar field would be linear in $x$,
\begin{equation}
\varphi\simeq\tilde{\varphi}+\varepsilon\sqrt{6}x,
\end{equation}
where $\tilde{\varphi}$ is a constant of integration. This gives an exponential potential (by setting $\tilde{\varphi}$ to zero) 
\begin{equation}
V(\varphi) = V_{0} \exp(-\varepsilon\sqrt{3/2}~\lambda~\varphi ).
\end{equation}
Using Eq. (\ref{ch1})  the \emph{n--field} $\phi$ is expressed in terms of the scalar field $\varphi$ as
\begin{equation}
\phi=  \phi_{0}\exp(\varepsilon \sqrt{3/2}~ \varphi)
\end{equation}
or, inversely, 
\begin{equation}
\varphi = \varepsilon \sqrt{2/3}~ \ln \bigg(\dfrac{\phi}{\phi_{0}}\bigg),
\end{equation}
where 
\begin{equation}
\phi_{0}=\dfrac{\sqrt{2}}{3H_{0}\sqrt{1-\Omega^{0}_{m}-v (1-\lambda/2)^{-1}}}.
\end{equation}
We may also recast the potential in terms of the \emph{n--field} according to
\begin{equation}
V(\phi) \propto \bigg(\dfrac{\phi_{0}}{\phi}\bigg)^{\lambda}.
\end{equation}

\subsection{Universe dominated by matter and potential terms}

At the present stage when the universe is undergoing an accelerated expansion and the scalar field has a \emph{negligible} velocity (its  kinetic term is negligible, compared to matter and potential terms), the scalar field reads
\begin{equation}
\label{key}
\varphi\simeq \bar{\varphi}+\dfrac{2\varepsilon\sqrt{3}}{3}\dfrac{\sqrt{\lambda}}{1-\lambda}~\mbox{arcsinh}\bigg[\sqrt{2v(\Omega_{m}^{0}(2-\lambda))^{-1}}~\exp\bigg (\frac{3}{2}(1-\lambda)x\bigg) \bigg]
\end{equation}
with a constant of integration $\bar{\varphi}$ and $\lambda \ne 1$. Eliminating then $x$ between (\ref{2}) and (\ref{key}) in favor of the scalar field $\varphi$, the potential $V(\varphi)$ is obtained in the form
\begin{equation}
V(\varphi)=V_{0}\bigg[\dfrac{\Omega_{m}^{0}(2-\lambda)}{2v}~\sinh^{2}\bigg(\frac{\sqrt{3}}{2}\frac{(1-\lambda)}{\sqrt{\lambda}}~\varphi\bigg) \bigg] ^{-\lambda/(1-\lambda)},
\end{equation}
where we make the change $\varphi-\bar{\varphi} \to \varphi$ for simplicity. Moreover, omitting the kinetic term in the scalar--field energy density the \emph{n-field} can be computed from Eq. (\ref{ch1}) as
\begin{equation}
\label{key1}
\phi=  \tilde{\phi}_{\lambda} \cosh\bigg(\frac{\sqrt{3}}{2}\frac{(1-\lambda)}{\sqrt{\lambda}}~\varphi\bigg) R_{\lambda}(\varphi),
\end{equation}
where  
\begin{equation}
\label{key2}
\tilde{\phi}_{\lambda} = \dfrac{\varepsilon (-1)^{(1-2\lambda)/[2(1-\lambda)]}~ \sqrt{2\lambda(2-\lambda)}}{3 H_{0}(1-\lambda)\sqrt{v}}
\bigg[\dfrac{\Omega_{m}^{0}(2-\lambda)}{2v} \bigg] ^{\lambda/[2(1-\lambda)]}
\end{equation}
and 
\begin{equation}
R_{\lambda}(\varphi) = \mbox{Hypergeometric2F1}\bigg[\dfrac{1}{2}, \dfrac{1}{2}\bigg(1-\dfrac{\lambda}{1-\lambda}\bigg), \dfrac{3}{2}, \cosh^{2}\bigg(\frac{\sqrt{3}}{2}\frac{(1-\lambda)}{\sqrt{\lambda}}~\varphi\bigg) \bigg].
\end{equation}
In this case, it is impossible to express the field $\varphi$ as a function of the \emph{n--field} $\phi$. However, taking particularly $\lambda=1/2$ the function $R_{\frac{1}{2}}(\varphi) = 1$ and  Eq. (\ref{key1}) takes the simple form
\begin{equation}
\phi=  \tilde{\phi}_{\frac{1}{2}}\cosh\bigg(\frac{1}{2}\sqrt{\dfrac{3}{2}}~\varphi\bigg), 
\end{equation}
what yields 
\begin{equation}
\varphi= 2 \sqrt{\dfrac{2}{3}}~ \mbox{arccosh}\bigg(\dfrac{\phi}{\tilde{\phi}_{\frac{1}{2}}}\bigg).
\end{equation}
Meanwhile, the potential $V(\phi)$ is given by
\begin{equation}
\label{key3}
V(\phi)= \beta \bigg[\bigg(\dfrac{\phi}{\tilde{\phi}_{\frac{1}{2}}}\bigg)^{2}-1\bigg]^{-1},
\end{equation}
where we assume $(\phi/\tilde{\phi}_{\frac{1}{2}})^{2} > 1$ to have a positive potential. For $(\phi/\tilde{\phi}_{\frac{1}{2}})^{2} \gg 1$, (\ref{key3}) reduces to 
\begin{equation}
V(\phi)\simeq \beta \bigg(\dfrac{\tilde{\phi}_{\frac{1}{2}}}{\phi}\bigg)^{2} \qquad \mbox{with} \qquad \beta = \dfrac{4 v V_{0}}{3\Omega_{m}^{0}}.
\end{equation}

\section{Conclusions}
We constructed a quintessence scalar field cosmology by introducing some quantities that significantly facilitated the resolution of the Friedmann equations. Considering a potential, that decreases exponentially  with e-folding number, we solved analytically the Friedmann equations  and then showed that the current cosmic acceleration is possible for a large value of the present potential, that is, for a small velocity of the scalar field. It was shown furthermore that the observed acceleration could not end up but last forever in the future. Finally we have reconstructed $V(\varphi)$ and $V(\phi)$ for two different epochs (early and recent times) of the universe from the expressions of $\varphi(x)$ and $\varphi(\phi)$.


\providecommand{\href}[2]{#2}\begingroup\raggedright\endgroup


\begin{thebibliography}{10}

\bibitem{Riess}A. G. Riess et al., Astron. J. \textbf{116}, (1998) 1009.

\bibitem{Perlmutter}S. Perlmutter et al., Astrophys. J. \textbf{517}, (1999) 565--586. 

\bibitem{Tegmark}M. Tegmark et al., Phys. Rev. D \textbf{69} (2004) 103501. 

\bibitem{Seljak}U. Seljak et al., Phys. Rev. D \textbf{71} (2005) 103515.

\bibitem{Adelman}J. K. Adelman--McCarthy et al., Astrophys. J. Suppl. \textbf{162} (2006) 38.

\bibitem{Abazajian}K. Abazajian et al., Astron. J. \textbf{126} (2003) 2081. 

\bibitem{Spergel1}D. N. Spergel et al., Astrophys. J. Suppl. \textbf{170} (2007) 377.

\bibitem{Page}L. Page et al., Astrophys. J. Suppl. \textbf{170} (2007) 335.

\bibitem{Hinshaw}G. Hinshaw et al., Astrophys. J. Suppl. \textbf{170} (2007) 288.

\bibitem{Jarosik}N. Jarosik et al., Astrophys. J. Suppl. \textbf{170} (2007) 263.

\bibitem{Weinberg}S. Weinberg, Rev. Mod. Phys., Vol. 61, No. 1, January 1989.

\bibitem{Amendola1}L. Amendola and S. Tsujikawa, CUP, 2010.

\bibitem{Ratra}B. Ratra and P. J. E. Peebles, Phys. Rev. D \textbf{37}, 3406 (1988).

\bibitem{Ratra1}P. J. E. Peebles and B. Ratra, ApJL \textbf{325}, L17 (1988).

\bibitem{Zlatev}I. Zlatev, L. Wang and P. J. Steinhardt, Phys. Rev. Lett. \textbf{82}, 896--899 (1999).

\bibitem{Caldwell}R. R. Caldwell, R. Dave and P. J. Steinhardt, Phys. Rev. Lett. \textbf{80}, 1582--1585 (1998).

\bibitem{Wetterich}C. Wetterich, Nuc. Phys. B \textbf{302},  668--696 (1988).

\bibitem{Wetterich1}C. Wetterich, Astron. Astrophys. \textbf{301}, 321--328 (1995).

\bibitem{Coble}K. Coble, S. Dodelson  and  J. Frieman, Phys. Rev. D \textbf{55}, 1851--1859 (1997).

\bibitem{Turner}M. S. Turner and M. White, Phys. Rev. D \textbf{56}, 4439--4443 (1997).

\bibitem{Boyle}L. A. Boyle, R. R. Caldwell and M. Kamionkowski, Phys. Lett. B \textbf{545}, 17--22 (2002).

\bibitem{Saridakis}Y.-F. Cai, E. N. Saridakis, M. R. Setare and J.-Q. Xia; Phys. Rept. 493, 1 (2010); E. N. Saridakis and S. V. Sushkov, Phys. Rev. D \textbf{81} (2010) 083510.


\bibitem{Copeland}E. J. Copeland, M. Sami and S. Tsujikawa, Int. J. Mod.Phys. D  \textbf{15}, 1753 (2006).

\bibitem{Copeland1}E. J. Copeland, A. R. Liddle and D. Wands, Phys. Rev. D \textbf{57}, 4686 (1998).

\bibitem{Ng}S. C. C. Ng, N. J. Nunes and F. Rosati, Phys. Rev. D \textbf{64}, 083510 (2001).

\bibitem{Varun}V. Sahni and A. Starobinsky, International Journal of Modern Physics D Vol. 15, No. 12 (2006) 2105-2132.

\bibitem{Padmanabhan}T. Padmanabhan, Phys. Rev. D \textbf{66}, 021301(R) (2002).

\bibitem{Bagla} J. S. Bagla, H. K. Jassal, and T. Padmanabhan, Phys. Rev. D \textbf{67}, 063504 (2003).











\end{thebibliography}
\end{document}